# Evidence of environmental strains on charge injection in silole based organic light emitting diodes.

N. Huby and L. Hirsch [*]

Université de Bordeaux ; CNRS ; UMR 5218 ; 351 cours de la Libération, 33405 Talence Cedex, France

L. Aubouy and Ph. Gerbier

Institut Charles Gerhardt ; UMR 5253, Université Montpellier II, CC 007, Place Eugène Bataillon, 34095 Montpellier Cedex, France

A. van der Lee

Institut Européen des Membranes - IEM (CNRS UMR 1919), Route de Mende, 34293 Montpellier Cedex 5, France

F. Amy and A. Kahn

Department of Electrical Engineering, Princeton University, Princeton, New Jersey 08544, USA

---

[*] Corresponding Author : lionel.hirsch@ims-bordeaux.fr

**ABSTRACT**

Using density functional theory (DFT) computations, we have demonstrated a substantial skeletal relaxation when the structure of 2,5-[bis-(4-anthracene-9-yl-phenyl]-1,1-dimethyl-3,4-diphenyl-silole (BAS) is optimized in the gas-phase comparing with the molecular structure determined from monocrystal X-ray diffraction. The origin of such a relaxation is explained by a strong environmental strains induced by the presence of anthracene entities. Moreover, the estimation of the frontier orbital levels showed that this structural relaxation affects mainly the LUMO that is lowered of 190 meV in the gas-phase. To check if these theoretical findings would be confirmed for thin films of BAS, we turned to ultraviolet photoemission spectroscopy / inverse photoemission spectroscopy and electro-optical measurements. Interestingly, the study of the current density / voltage and luminance / voltage characteristics of an ITO/PEDOT/BAS/Au device clearly demonstrated a very unusual temperature-dependant behavior. Using a thermally-assisted tunnel transfer model, we found that this behavior likely originated from the variation of the electronic affinity of the silole derivative with the temperature. The thermal agitation relaxes the molecular strains in thin films as it is shown when passing from the crystalline to the gas-phase. The relaxation of the intramolecular thus induces an increase of the electronic affinity and, as a consequence, the more efficient electron injection in organic light-emitting diodes.

PACS: 72.80.Le, 73.50.-h, 71.15.Mb, 73.40.Sx

# 1  Introduction

The wide range of organic semiconductors used as emissive layers in organic light-emitting diodes (OLED) for displays or lighting applications implies specific investigations of each molecular family. Understanding phenomena related to the molecular properties is crucial to these applications, and usually involves fundamental investigations. Conjugated molecules in solid state are weakly bond by van der Waals interactions or hydrogen bonds. Although the binding energy is low, the molecular environment can play a fundamental role when it induces steric strains [1]. Bond strain is accompanied by a loss of molecular symmetry, which induces an unusual electron distribution and a variation of the band gap.

The investigations presented in this paper are carried out with a new silole derivative: 2,5-[bis-(4-anthracene-9-yl-phenyl]-1,1-dimethyl-3,4-diphenyl-silole, herein named BAS (see Figure 1). Silole-based compounds have been proposed as a new class of electroluminescent materials [2,3]. They are mainly electron transport semiconductors [4], and have a low-lying lowest unoccupied molecular orbital (LUMO) due to an effective interaction between the $\sigma^*$-orbital of the silicon atom and the $\pi^*$-orbital of the butadiene fragment [5,6]. Two anthracene groups, well-known as hole-transport groups in molecular films [7,8], are attached to the central silole ring. By combining the electron-transport properties of the silole and the hole transport ability of the anthracene, the intra-molecular balance of charge carriers should be improved in single-layer devices [9].

Density Functional Theory (DFT) computations and crystallographic structure analyses were carried out. Geometrical parameters of BAS have been measured in the single crystal and compared to the ones obtained from the optimization of the molecular geometry in the gas-phase. In contrast with what we had previously observed with others silole derivatives using the same methodology, some significant discrepancies are found between the as-found and the optimized structures [9-13]. The origin of these discrepancies may be explained by a relaxation phenomenon reflecting strong environmental strains in the crystal lattice. This observation prompted us to look at the role of supramolecular environmental strain and its consequences on electron injection in a layer of silole-derivative semiconductor. Indeed, the balance of charge carriers is a fundamental parameter to optimize the performances of emissive layers.

Electro-optical study, Ultraviolet Photoemission Spectroscopy / Inverse Photoemission Spectroscopy (UPS/IPES) measurements are carried out to investigate the relaxation of environmental strains on BAS molecules in the solid state with increasing the temperature. The consequence of that relaxation results in an increase of the electron affinity (*EA*), which plays a fundamental role in the electron injection probability.

## 2  Results and discussion

### 2.1  Structural characterization

BAS was synthesized by an adaptation of the general procedure reported by Tamao et al. involving the intramolecular reductive cyclization of dimethyl-bis(phenylethynyl)silane followed by the coupling reaction with 9-(4-bromophenyl)anthracene [14]. BAS crystallizes in the P$_{-1}$ triclinic space group with a = 10.094(2) Å, b = 10.998(2) Å, c = 18.808(2) Å, α = 85.90(1) °, β = 84.34(1) °, γ = 88.44(2)° and Z = 2. The diffraction intensities for BAS were collected at the joint X-ray Scattering Service of the Institut Européen des Membranes and the Institut Charles Gerhardt of the Université de Montpellier II, France, at 175 K using an Oxford Diffraction Xcalibur I diffractometer and Mo-*Kα* photons (*λ*=0.71073 Å). The structure of BAS, $C_{58}H_{42}Si$, was solved by direct methods using SIR2002 [15] and refined by least-squares methods using CRYSTALS [16,17]. As shown in figure 2, the molecule displays a propeller-like arrangement of the four benzene rings as usually observed with other tetra-arylsiloles. The dihedral angles (ϕ) between the mean planes of the central silole (S, see Figure 1 for labeling of the rings), the 2,5-benzene rings (P$_1$ and P$_2$) and the anthracenyl groups ($A_1$ and $A_2$) are given in Table 1. While the first three values fall within the range of those generally reported for related structural units, it is worthy of note that the forth one ($\phi(P_2 - A_2)$ = 64.2(3)°), is substantially lower than the smallest value encountered for 9-arylanthracene derivatives (*ca.* 70°), and induces appreciable intramolecular steric interactions. Significant π-π stacking interactions are found between $A_1$ anthracenyl groups of adjacent BAS molecules that in fact determine the cohesion of the three-dimensional intermolecular structure (Figure 3). The lowest distance between two phenyl centroids is 3.711(4) Å with 23.20° and 20.81° for the angles between the ring normals and the centroid vectors. These values are in accordance within the accepted range of values for appreciable π-π interactions [18].

The geometry optimization of BAS in the gas phase was carried out using density functional theory (DFT) calculations with the B3LYP functional. Due to the size of the molecule, geometry optimization without symmetry constrains was performed with the 6-31G basis set to the standard convergence criteria as implemented in Gaussian G03w software [19]. In the gas phase, the geometry of BAS approaches the $C_2$ symmetry by relaxing the geometrical constraints between the different rings in order to minimize the steric interactions and to increase the conjugation throughout the π-system. Thus, the values of the torsion angles between the silole and the adjacent phenyl rings are decreased (Table 1) whereas the values of the torsion angles between the phenyl rings and the anthracenyl groups are increased. Among the siloles we have already studied these results are quite unusual (see above). For instance, in the case of 1,1-dimethyl-2,5-bis(p-2,2'-dipyridylaminophenyl)-3,4-diphenylsilole (DMPPS, Figure 1) for which we have already described the mechanism of injection, both the crystal and the optimized structures are very similar [9,20,21]. In the DMPPS molecule, the anthracene side-groups have been replaced by dipyridylaminophenyl groups, and for that reason, no structural constraint is observed in the solid state. Therefore, it is likely that the propensity of anthracene to adopt a π-stacking structure induces sufficient environmental strain to bring about the BAS molecules to adopt such an energetically costly conformation in the crystal lattice (see Figure 3).

## 2.2 Electronic level characterization

Ultraviolet photoemission spectroscopy was carried out using He I (21.2 eV) and He II (40.8 eV) sources in conjunction with a double-pass cylindrical mirror analyzer. The resolution of the photoemission system was 150 meV as determined from the width of the Fermi edge on an electrode of freshly deposited Au. Inverse photoemission spectroscopy was carried out using an electron energy scan from 5 to 15 eV with a system resolution of 500 meV. UPS and IPES measurements (Figure 4) were carried out on an 8 nm BAS vapor-deposited thin film on a gold substrate to estimate the energies of its filled and empty orbitals. The measured ionization potential (*IP*) and *EA*, defined as the energy difference between vacuum level and onset of occupied and unoccupied states, are 6 eV and 1.69 eV, respectively. The single particle energy gap between the onsets of LUMO and HOMO equals 4.31 eV. To complete this study, DFT single point calculations were performed on both the X-ray and optimized geometries of BAS using the B3LYP functional with a 6-31G* split valence plus polarization basis set. The calculated HOMOs and LUMOs (Figure 5) do not much vary in appearance. The major changes in the electronic distribution reflect the symmetrization of the geometry when going from the solid to the gas phase (optimized geometry). On the other hand, the variation of the geometry of BAS from gas-phase to solid (crystalline) phase has a strong influence on the energy parameters, in particular the electron affinity (Table 2). It is noteworthy that the HOMO level is not much affected from the solid state to the relaxed gas-phase, contrary to the LUMO level. The latter is clearly stabilized in the relaxed gas-phase. The *EA* increases from -1.81 eV in the solid state to -2.00 eV in the gas-phase. Simultaneously, the *IP* varies only slightly, from -5.37 eV in the solid state to -5.39 eV in the gas phase.

## 2.3 Thermally assisted electron tunneling injection

Thus, structural modifications induce significant variation of the electronic structure of the molecule. At low temperature (90 K), the solid state is frozen. When the temperature increases, thermal agitation should relax the environmental strains and thus increase *EA*. To further probe this important mechanism, we built a device based on the following structure: ITO/PEDOT/BAS (80nm)/Au. The gold cathode is chosen for its high work function in order to be sure to characterize the BAS/Au interface. Variation in *EA* with temperature should have a strong effect on the electron injection at this interface. Details of device manufacturing can be found in reference [20].

Typically, transport over a huge barrier is analyzed in terms of tunnel injection [22]. Figure 6 shows the current density-voltage (J-V) and luminance-voltage (L-V) characteristics for temperatures ranging from 90 K to 235 K carried out in a liquid nitrogen cryostat chamber under rotary pump vacuum ($10^{-3}$ mbar) and recorded with a Semiconductor Characterization System *Keithley* 4200 with a monitored photodiode. The first bias window (Domain I) is associated with hole-only injection and transport, because electron injection is ruled out by the large Au-LUMO potential barrier. The energy barrier between the onset of the empty states and the Au Fermi level is 2.64 eV (Figure 4). No luminescence is detected in this regime. The luminance appears in the second bias window (Domain II), which is the sign of significant injection of both types of charge carriers. The associated current in Domain II is much more dependent on the applied voltage than in Domain I, indicating that it is dominated by electrons. Considering the 2.64 eV electron injection barrier at the BAS/Au interface (see Figure 4), thermionic injection is not possible and only tunnel injection at high electric field needs to be considered. In the case of DMPPS, J-V and L-V characteristics were temperature-independent and described using the Fowler-Nordheim tunnel transfer model [20]. On the contrary, in the case of BAS, J-V as well as L-V curves are thermally activated (Figure 6). The origin of this phenomenon is likely due to the relaxation of the environmental strains when increasing the temperature i.e. the higher the temperature, the lower the environmental strain pressure. DFT calculations summarized in Table 2 highlight that the LUMO and HOMO are strongly influenced by the torsion angles. Raghunath et al. recently obtained similar results in anthracene based molecule [8]. In the following model, we will consider that the environmental strain relaxation is Maxwell-Boltzmann temperature dependent.

We now introduce a thermally assisted tunnel injection model across a triangular potential barrier $\psi(x)$. The Fermi level is taken as the reference level. The potential barrier thickness is $a$, with $\psi(a) = 0$. The potential barrier height at the interface, i.e. LUMO - $E_F$(Au), is $\psi(0) = \phi_c$. Both $a$ and $\psi(0)$ are temperature dependant. $\psi(x)$ and $a$ are defined in equations 1 and 2:

$$\psi(x) = \phi_c \left(1 - \exp\left(-\frac{W}{k_B T}\right)\right) - Fx \tag{1}$$

$$a = \frac{\phi_c}{F}\left(1 - \exp\left(-\frac{W}{k_B T}\right)\right) \tag{2}$$

where $F = V_{ap} / d$ is the electric field, $V_{ap}$ the applied voltage, $d$ the thickness of the BAS layer, $\phi_c = \phi_m$ (Au) – $EA$ (BAS) = $\psi(0)$, $W$ is the activation energy to relax the molecule of its environmental strain, $E$ is the energy of the electrons and $k_B = 1.38 \times 10^{-23}$ J/K is the Boltzmann's constant. A gold cathode was chosen to have a high barrier $\phi_c$. Thus, the current density is assumed to be directly proportional to the tunnel transfer probability $P$ [23,24] expressed by the equation (3) i.e. $J = K \times P$:

$$P = \exp\left(-2\int_0^a \frac{2m}{\hbar^2}(\psi(x) - E)^{1/2} dx\right) \tag{3}$$

Incorporating $\psi(x)$, $\psi(0)$ and $a$ into equation (3) yields after integration:

$$P = \exp\left(\frac{-4(mq)^{1/2}E^{3/2}\sqrt{2}d}{3\hbar V}\right)\exp\left(\frac{-4(mq)^{1/2}\sqrt{2}d}{3\hbar V}\left(-E+\phi_c\left(1-\exp\left(-\frac{qW}{k_BT}\right)\right)\right)^{3/2}\right) \quad (4)$$

Where $m = 9.1\times10^{-31}$ kg and $q = 1.6\times10^{-19}$ C being the mass and the charge of the electron respectively, $d = 80\times10^{-9}$ m is the thickness of the BAS layer and $\hbar = h/2\pi = 1.05\times10^{-34}$ J.s the reduced Planck constant.

In order to analytically solve equation (4), we have supposed that $E \to 0$ *i.e.* electrons are supposed to be close to the Fermi level. This approximation is supported by two reasons: *i)* the probability to find electrons 100 meV over the Fermi level at $T < 250$ K is very low (less than 1%), and *ii)* if thermally dependent tunneling transport was resulting from the Fermi-Dirac statistic occupation of electrons over $E_F$ with increasing the temperature, such phenomenon should occur whatever the material. Actually, in that case of DMPPS-based devices, no thermally dependent tunneling transport was observed [20]. As a result, thermally effect originating from the occupation of electrons over the Fermi level in the device with a gold cathode can be neglected. Equally, we neglect any variation of the interfacial dipole with the temperature.

Consequently, substituting the numerical values in equation (4) yields:

$$P \approx \exp\left(\frac{546}{V}\left(\phi_c\left(1-\exp\left(-11601\times\frac{W}{T}\right)\right)\right)\right) \quad (5)$$

Figure 7 shows the J-V characteristics and the corresponding Arrhenius plots for Domain II in Figure 6. The experimental data are represented by the full symbols. The theoretical values (full curves) have been calculated from equation (5) using $W = 35$ meV and $\phi_c = 1.3$ eV. A thermal energy of 35 meV corresponds to a temperature of about 400 K, which is almost the BAS melting point. Above the fusion temperature, molecular environmental strain is inexistent. The weak value of $\phi_c$ compared to the difference $\phi_m$ (Au) – $EA$ (BAS) indicates the presence of an electrical dipole at the interface BAS/Au [25,26] and/or a possible diffusion of gold atoms in the BAS layer. The model of thermally assisted tunnel transfer fits experimental data. The higher the temperature, the lower the environmental strain on the BAS molecule. The consequence is an increase in *EA* and an enhancement of electron injection in BAS-based devices. The effective variation of *EA* can be calculated from equation (1) using the fitting parameters. The variation of *EA* from 90 to 235 K is around 0.17 eV which is almost in accordance with that calculated with DFT.

Optical absorption or photoluminescence or electroluminescence (EL) does not directly provide data on LUMO and HOMO levels because in organic compounds, exciton binding energy is roughly 1 eV. However, a red-shift should be observed in the EL spectra by increasing the temperature of the sample. As a consequence, we recorded EL spectra for temperature varying from 83 K to 297 K on ITO/PEDOT-PSS/BAS (80 nm)/Ca devices. Electroluminescence spectra were collected using an optic fiber connected to an *OceanOptics HR2000* spectrometer while OLEDs were submitted to a constant current density of 10 mA.cm$^{-2}$. OLEDs were biased for less than one second to reduce any heating effect during the measurement. Results are presented in figure 8. A slight red-shift, of almost 70 meV, is observed when the temperature is increased from 83 K to 297 K. Actually, two phenomena are in competition: a red-shift resulting from the stabilization of the LUMO and a blue-shift (commonly observed in polymers and small molecules) due to a statistical occupation of higher excited states at high temperature [27-29]. Thus, this observed red-shift qualitatively confirm the relaxation of environmental strains with increasing the temperature.

## 3  Conclusion

In summary, we have demonstrated that the π-stacking of anthracene entities induces environmental strains on the BAS molecules affecting the molecular electronic levels. DFT calculations are supported by electrical measurements and by electro-optical study. A thermally assisted tunnel transfer was developed. All data are consistent with a modification of the *EA* with the temperature. Environmental strains on BAS molecule relax when increasing the temperature. It means that the energy optimization of the solid state induce torsion angles which put the BAS molecule of its stable energy configuration. This phenomenon is highlighted by the increase of the electron affinity and an enhancement of the electron injection at the interface Organic/cathode.

**Acknowledgments :**

This work was partially supported by the "Région Aquitaine", the "Région Languedoc-Roussillon" and the STII Department of French CNRS, and the National Science Foundation (DMR-0408589).

**REFERENCES**

1   M. Senna, J. Mat. Sci. **39,** 4995 (2004).
2   L. C. Palilis, A. J. Mäkinen, M. Uchida, and Z. H. Kafafi, Appl. Phys. Lett. **82,** 2209 (2003).
3   H. Murata, Z. H. Kafafi, and M. Uchida, Appl. Phys. Lett. **80,** 189 (2002).
4   H. Murata, G. G. Malliaras, M. Uchidac, Y. Shenb, and Z. H. Kafafi, Chem. Phys. Lett. **339,** 161 (2001).
5   S. Yamaguchi and K. Tamao, Bull. Chem. Soc. Jpn **69,** 2327 (1996).
6   S. Yamaguchi and K. Tamao, in *Chemistry of Organic Silicon Compounds*; Vol. 3, edited by Z. Rappoport and Y. Apeloig (John Wiley & Sons, Chichester, 2001), p. 641.
7   K. Danel, T. H. Huang, J. T. Lin, Y. T. Tao, and C. H. Chuen, Chem. Mat. **14,** 3860 (2002).
8   P. Raghunath, M. A. Reddy, C. Gouri, K. Bhanuprakash, and V. J. Rao, J. Phys. Chem. A **110,** 1152 (2006).
9   L. Aubouy, P. Gerbier, N. Huby, G. Wantz, L. Vignau, L. Hirsch, and J. M. Janot, New J. Chem. **28,** 1086 (2004).
10  J. Lee, Q.-D. Liu, M. Motala, J. Dane, J. Gao, Y. Kang, and S. Wang, Chem. Mat. **116,** 1869 (2004).
11  N. Roques, P. Gerbier, U. Schatzschneider, J. P. Sutter, P. Guionneau, J. Vidal-Gancedo, J. Veciana, E. Rentschler, and C. Guérin, Chem. Eur. J. **12,** 5547 (2006).
12  N. Roques, P. Gerbier, Y. Teki, S. Choua, P. Lesniakovà, J. P. Sutter, P. Guionneau, and C. Guérin, New J. Chem. **30,** 1319 (2006).
13  F. Habrard, T. Ouisse, O. Stéphan, L. Aubouy, P. Gerbier, L. Hirsch, N. Huby, and A. v. d. Lee, Synth. Met. **156,** 1262 (2006).
14  K. Tamao, S. Yamagushi, M. Shiozaki, Y. Nakagawa, and Y. Ito, J. Am. Chem. Soc. **114,** 5867 (1992).
15  M. C. Burla, M. Camalli, B. Carrozzini, G. L. Cascarano, C. Giacovazzo, G. Polidori, and R. Spagna, J. Appl. Cryst. **36,** 1103 (2003).
16  P. W. Betteridge, J. R. Carruthers, R. I. Cooper, K. Prout, and D. J. Watkin, J. Appl. Cryst. **36,** 1487 (2003).
17  1409 [$I>2\sigma(I)$] data and 532 parameters leading to the usual crystallographic reliability factors $R_1$=0.0262 and $wR_2$=0.0250 have been used for CRYSTAL. The H atoms were initially refined with soft restraints on the bond lengths and angles to regularize their geometry with the C---H distance in the range 0.93-0.98 Å and $U_{iso}$(H) in the range 1.2-1.5 times $U_{eq}$ of the parent atom, after which the positions were refined with riding constraints. CCDC-633105 contains the supplementary crystallographic data for this paper. These data can be obtained free of charge from the Cambridge Crystallographic Data Centre via http://www.ccdc.cam.ac.uk/products/csd/request/.).
18  C. Janiak, J. Chem. Soc., Dalton Trans.**,** 3885 (2000).
19  M. J. Frisch, G. W. Trucks, H. B. Schlegel, G. E. Scuseria, M. A. Robb, J. R. Cheeseman, J. A. Montgomery, T. Vreven, K. N. Kudin, J. C. Burant, J. M. Millam, S. S. Iyengar, J. Tomasi, V. Barone, B. Mennucci, M. Cossi, G. Scalmani, N. Rega, G. A. Petersson, H. Nakatsuji, M. Hada, M. Ehara, K. Toyota, R. Fukuda, J. Hasegawa, M. Ishida, T. Nakajima, Y. Honda, O. Kitao, H. Nakai, M. Klene, X. Li, J. E. Knox, H. P. Hratchian, J. B. Cross, V. Bakken, C. Adamo, J. Jaramillo, R. Gomperts, R. E. Stratmann, O. Yazyev, A. J. Austin, R. Cammi, C. Pomelli, J. W. Ochterski, P. Y. Ayala, K. Morokuma, G. A. Voth, P. Salvador, J. J. Dannenberg, V. G. Zakrzewski, S. Dapprich, A. D. Daniels, M. C. Strain, O. Farkas, D. K. Malick, A. D. Rabuck, K.

Raghavachari, J. B. Foresman, J. V. Ortiz, Q. Cui, A. G. Baboul, S. Clifford, J. Cioslowski, B. B. Stefanov, G. Liu, A. Liashenko, P. Piskorz, I. Komaromi, R. L. Martin, D. J. Fox, T. Keith, M. A. Al-Laham, C. Y. Peng, A. Nanayakkara, M. Challacombe, P. M. W. Gill, B. Johnson, W. Chen, M. W. Wong, C. Gonzalez, and J. A. Pople, 03, revision C.02 ed. (Gaussian, Inc., Wallingford, CT, 2004).

[20] N. Huby, L. Hirsch, G. Wantz, L. Vignau, A. S. Barrière, J. P. Parneix, L. Aubouy, and P. Gerbier, J. Appl. Phys. **99,** 084907 (2006).

[21] J. Lee, Q.-D. Liu, D.-R. Bai, Y. Kang, Y. Tao, and S. Wang, Organometallics **23,** 6205 (2004).

[22] I. D. Parker, J. Appl. Phys. **75,** 1656 (1994).

[23] J. G. Simmons, Phys. Rev. **155,** 657 (1967).

[24] S. M. Sze, *Semiconductor devices* (Wiley Interscience, New York, 1981).

[25] I. G. Hill, A. Rajagopal, A. Kahn, and Y. Hu, Appl. Phys. Lett. **73,** 662 (1998).

[26] S. Narioka, H. Ishii, D. Yoshimura, M. Sei, Y. Ouchi, K. Seki, S. Hasegawa, T. Miyazaki, Y. Harima, and K. Yamashita, Appl. Phys. Lett. **67,** 1899 (1995).

[27] G. Wantz, L. Hirsch, N. Huby, L. Vignau, A. S. Barrière, and J. P. Parneix, J. of Appl. Phys. **97,** 034505 (2005).

[28] M. Anni, M. E. Caruso, S. Lattante, and R. Cingolani, J. Chem. Phys. **124,** 134707 (2006).

[29] G. Y. Zhong, X. M. Ding, J. Zhou, N. Jiang, W. Huang, and X. Y. Hou, Chem. Phys. Lett. **420,** 347 (2006).

**Table and figure captions**

**Table 1:** Dihedral angles ($\phi$) between the mean planes of the central silole (S), the 2,5-benzene rings ($P_1$ and $P_2$) and the anthracenyl groups ($A_1$ and $A_2$).

**Table 2:** Calculated HOMOs and LUMOs energy levels on both the X-ray and optimized geometries of BAS using the B3LYP functional with a 6-31G* split valence plus polarization basis set.

**Figure 1:** 2,5-[bis-(4-anthracene-9-yl-phenyl]-1,1-dimethyl-3,4-diphenyl-silole (BAS). S is the central silole ring, $P_1$, and $P_2$ the 2,5-benzene rings and $A_1$ and $A_2$ the anthracenyl groups, and 1,1-dimethyl-2,5-bis(p-2,2'-dipyridylaminophenyl)-3,4-diphenylsilole (DMPPS).

**Figure 2:** Three-dimensional structure of BAS such as found from the analysis of the X-ray data (see Figure 1 for the meaning of labels). The displacement ellipsoids are drawn at the 50% probability level. H atoms are shown as spheres of arbitrary radius.

**Figure 3:** $\pi$-$\pi$ aromatic interaction between anthracenyl moieties of adjacent BAS molecules.

**Figure 4:** UPS and IPES spectra performed on a BAS thin film (80 nm thick) deposited under ultra high vacuum on a clean gold substrate. The energy scale is referenced to the Fermi level. Values of HOMO, LUMO, *EA* and *IP* are reported in the figure.

**Figure 5:** DFT computed frontier orbitals of BAS in the crystal lattice and in the gas phase (optimized geometry).

**Figure 6:** Current density – voltage and luminance – voltage characteristics of the device ITO/PEDOT/BAS 80 nm/Au for temperature comprised between 90 K and 235 K.

**Figure 7:** Experimental (symbols) and theoretical (lines) J-V curves (a) and the corresponding Arrhenius plot (b).

**Figure 8:** Normalized electroluminescent spectra recorded at 87 K, 148 K and 297 K on ITO/PEDOT-PSS/BAS (80 nm)/Ca devices.

|  | $\phi(S - P_1)$ (°) | $\phi(S - P_2)$ (°) | $\phi(P_1 - A_1)$ (°) | $\phi(P_2 - A_2)$ (°) |
| --- | --- | --- | --- | --- |
| BAS crystals | 57.5(3) | 59.3(3) | 72.7(3) | 64.2(2) |
| BAS optimized | 49.1 | 49.1 | 77.7 | 77.6 |

Table 1    Huby et al.

|  | HOMO-1 (eV) | HOMO (eV) | LUMO (eV) | LUMO+1 (eV) |
| --- | --- | --- | --- | --- |
| BAS crystals | -5.43 | -5.37 | -1.81 | -1.76 |
| BAS optimized | -5.42 | -5.39 | -2.00 | -1.90 |

Table 2     Huby et al.

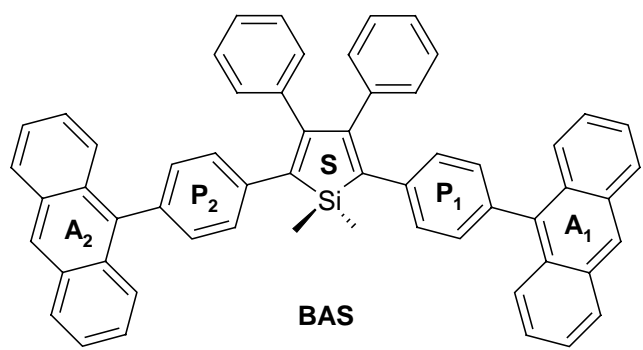

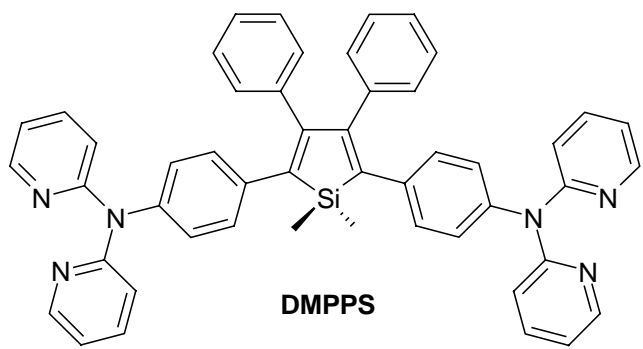

FIGURE 1     Huby et al.

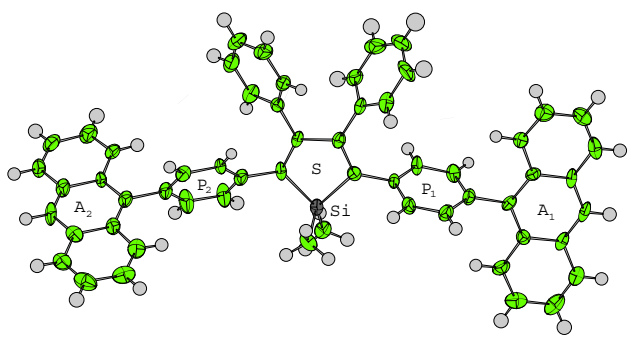

FIGURE 2    Huby et al.

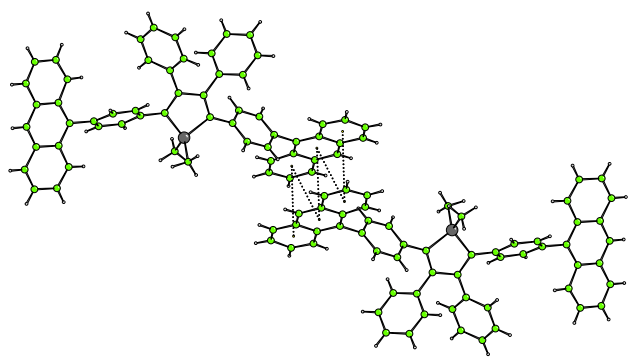

FIGURE 3      Huby et al.

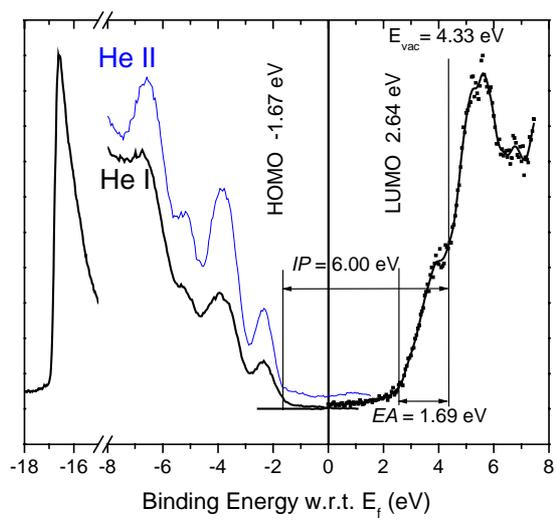

FIGURE 4    Huby et al.

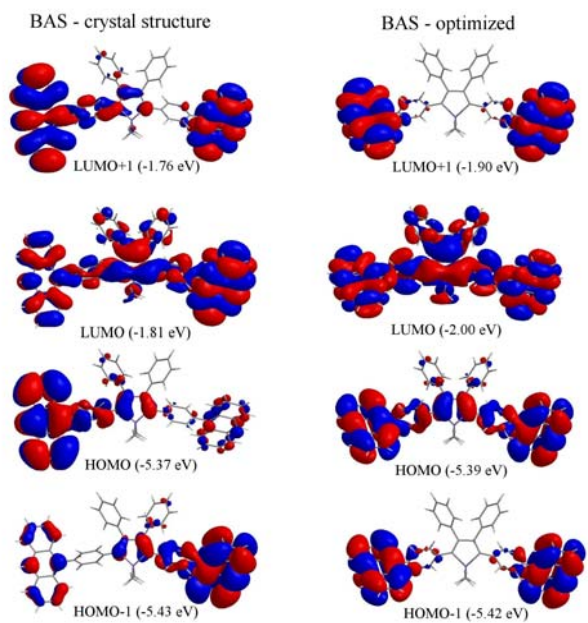

FIGURE 5    Huby et al.

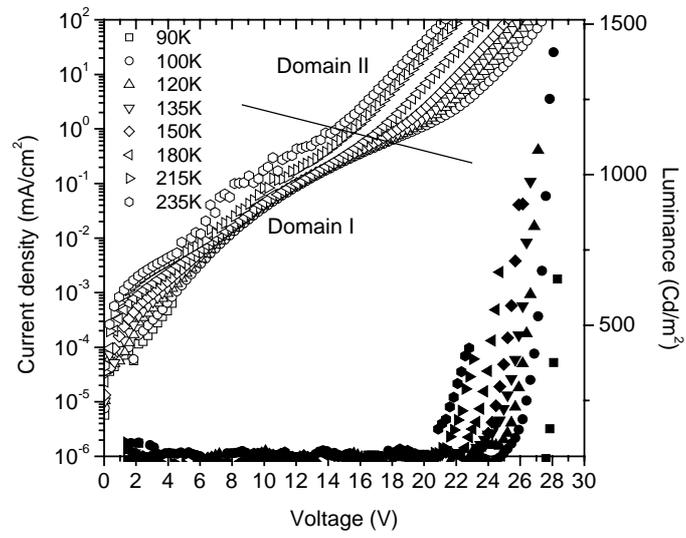

FIGURE 6　　Huby et al.

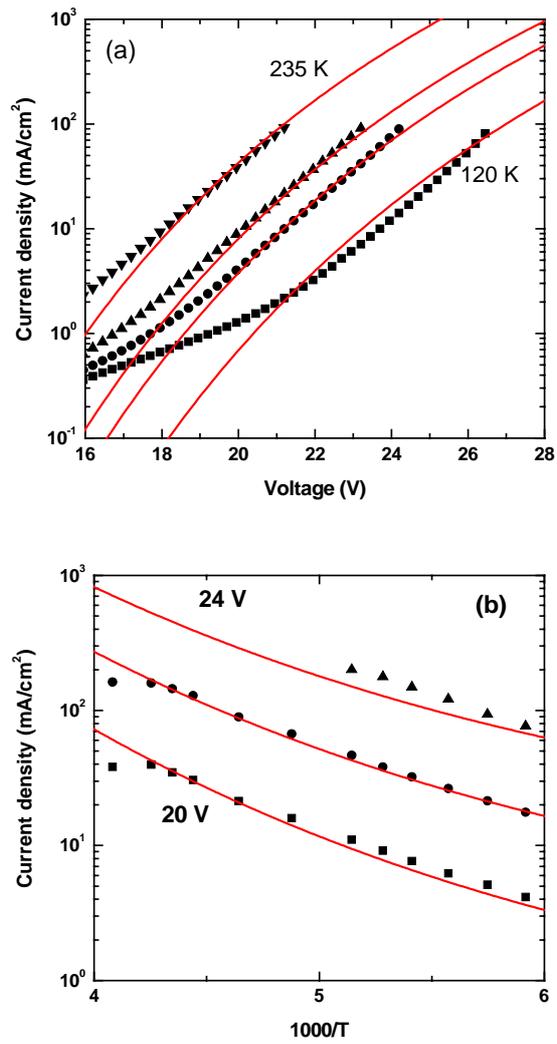

FIGURE 7    Huby et al.

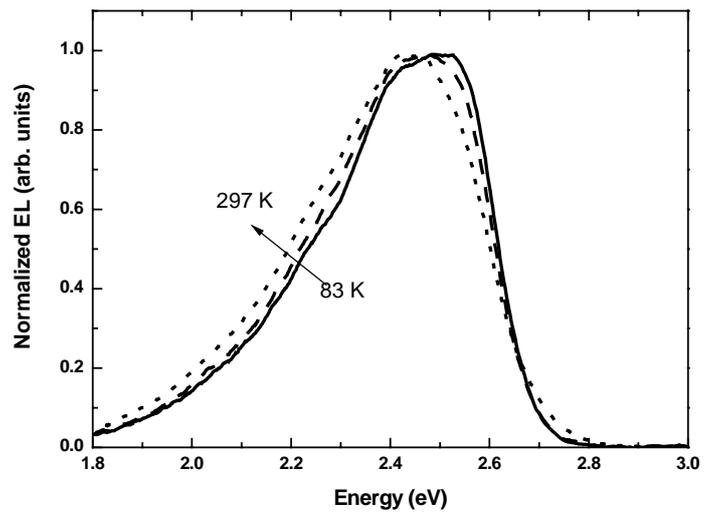

FIGURE 8    Huby et al.